\documentclass[galaxies,article,accept,oneauthor,pdftex]{mdpi}

\usepackage{xcolor}
\firstpage{1}
\pubvolume{9}
\issuenum{4}
\articlenumber{80}
\pubyear{2021}
\copyrightyear{2021}
\externaleditor{Academic Editors: Antara R. Basu-Zych, Dimitris M. Christodoulou and Phil Edwards} 
\datereceived{27 June 2021}
\dateaccepted{7 October 2021}
\datepublished{19 October 2021}
\hreflink{https://doi.org/\linebreak 10.3390/galaxies9040080} 

\newcommand{\aap}{Astron.  Astrophys.}

\newcommand{\apj}{Astrophys.  J.}

\newcommand{\nat}{Nature}

\newcommand{\prd}{Phys. Rev. D}

\newcommand{\apjs}{Astrophys. J. Suppl.}

\graphicspath{{./Figures/}}
\Title{Synthetic Neutrino Imaging of a Microquasar}

\TitleCitation{Synthetic Neutrino Imaging of a Microquasar}

\Author{{Theodoros }
Smponias}

\AuthorNames{Theodoros Smponias}

\AuthorCitation{Smponias, T.}

\address[1]{%
Directorate of Secondary Education of the Dodekanese, Zephyros, 85100 {Rhodes}, Greece; t.smponias@hushmail.com}

\abstract{Microquasar binary stellar systems emit electromagnetic radiation and high-energy particles over a broad energy spectrum. However, they are so far away that it is hard to observe their details. A simulation  offers the link between relatively scarce observational data and the rich theoretical background. In this work, high-energy particle emission from simulated twin microquasar jets is calculated in a unified manner. From the cascade of emission within an element of jet matter to the dynamic and radiative whole jet model, the series of physical processes involved are integrated together. A programme suite assembled around model data produces synthetic images and spectra directly comparable to potential observations by contemporary arrays. The model is capable of describing a multitude of system geometries, incorporating increasing levels of realism depending on  need and  available computational resources. As an application, the modelling process is applied to a typical microquasar, which is synthetically observed from  different angles using various imaging geometries. Furthermore, the resulting intensities are  comparable to the sensitivity of existing detectors. The combined background emission from a potential distribution of microquasars is also~modelled.}


\begin{document}

\section{Introduction}
\label{intro}

Microquasars (MQ) comprise a binary stellar system  where a main sequence star orbits a compact object, either a neutron star or a black hole \cite{Mirabel99}. Matter from the star accretes onto the collapsed stellar remnant, resulting in the production of twin relativistic jets pointing in opposite directions. Those jets emit over a broad spectrum, from radio to very high-energy (VHE) $\gamma$ rays and neutrinos \cite{Romero2003,Bednarek2005,Bosch_Ramon_2007,Reynoso2008,Reynoso2009,Christiansen2013high,Zhang2010,Reynoso2019}.

As mentioned in \cite{Romero2003}, apparent superluminal motion in certain MQs indicates the presence of bulk hadron flows in the jets. The assumption of equipartition \cite{Reynoso2009} leads to high magnetic field estimates for the jet \cite{Koessl1990}. This, coupled with the fluid approximation for the jet matter due to the presence of tangled magnetic fields \cite{Rieger2006,Rieger2019}, allows for  magnetohydrodynamic (MHD) approximation for the jets. A toroidal magnetic field component may retain jet collimation over considerable distances along its path \cite{Koessl1990,Singh2019MHD}. Moreover, external confinement from surrounding winds is equally important \cite{Hughes1991,Reynoso2009}.

In order to study the jets, a selection from among the wealth of theoretical results is compared to observations of those remote systems. The relative scarcity of detailed data is complemented by the use of numerical simulations of a jet system, where a model setup is evolved and then imaged synthetically. The final model emissions are placed next to observations, running many examples until a match is achieved. If no positive detections exist yet, then a general match to theoretical results and   the sensitivity of active observing arrays is desired. As a next step, going backwards, the jet model is reverse-engineered to its initial boundary and generally internal or unobservable conditions, which emerge as the link between jet theory and observations.

The above process can offer increased insight into the inner physical workings of the jets and their surroundings, allowing for their study as a complex, evolving dynamical system. A more accurate description of the system of interest is then obtained.

In this paper, the production of VHE neutrinos from generic MQ jets is modelled using the method of dynamic and radiative relativistic MHD simulation. A set of surrounding winds assists with the confinement of the jets, adding realism to the model.

Within the jets, a complex turbulent environment allows for the production of a variety of different signals, from radio to X and $\gamma$ rays. Furthermore, cascades of high-energy particles produced in the jets lead to an ecosystem of different particle populations connected through transport phenomena. The production of neutrinos that leave the system opens the possibility of detection on Earth from modern arrays.

The solution of the transport equation from one particle distribution to the next, along a cascade, allows for the expression of the intensity of emitted neutrinos as a function of dynamic and radiative jet parameters at a given point. This way, local model parameters at each space-time point in the model jet are directly connected to the final particle emission at the same point. Repeating the latter process for a number of energies provides a neutrino energy spectrum at each jet space-time point. Line-of-sight integration follows, leading to the production of a synthetic neutrino image of the system and a whole-jet neutrino energy~spectrum.

The paper is organized as follows. In Section \ref{methodology}, the theoretical background of the work is presented. In Section \ref{neutrinos}, the emission of particles from the jet is obtained. In Section \ref{results},  results are presented and discussed. Normalization and equipartition (and the synthetic imaging process) are described in Appendixes \ref{Normalization} and \ref{Equipartition_calculation} respectively.












\section{Theoretical Setup}
\label{methodology}

In our generic MQ model, an accretion disk is assumed around the compact object~\cite{Fabrika2004}.  Twin jets emanate from the vicinity of the collapsed star, collimated by a toroidal magnetic field component. Adopting a heavier pair of jets, their kinetic power was set to \mbox{$L_k$ = 2~$\times$~10$^{38}$~ergs$^{-1}$} (see Appendix \ref{Normalization}). The authors of \cite{Reynoso2009} argued a 10\% Eddington luminosity jet power, leading to  $L_k$~=~10$^{38}$ ergs$^{-1}$ for a 10 $M_{\odot}$ black hole, which is comparable to our simulation. Furthermore, {for the ratio of proton jet power $L_{p}$ to electron jet power $L_{e}$}, the same authors argue either $\frac{L_{p}}{L_{e}}\simeq 100$ or $\simeq$1; we  adopted the former hypothesis, favouring protons. As a first implementation, we  calculate neutrino emission originating from the influence of the high-energy proton distribution, while there is also a potential comparable contribution from the corresponding high-energy electron distribution \cite{Reynoso2009}.

In the jets, equipartition is assumed between kinetic {($\rho_k$)} and magnetic {($\rho_B$)} energy densities, meaning $\rho_k=\rho_B$; therefore,  {at each jet point ${\vec{r}} {z}$,}  the {CGS} magnetic field is \mbox{$B({\vec{r}} {z})=\sqrt{8 \pi \rho_{{\vec{r}} {z} } }$}~ \cite{Rieger2006,Rieger2019}, a close match with the \emph{B} used in the simulation (see Appendix \ref{Equipartition_calculation}). External magnetic fields tend to be quite smaller \cite{Kolo2017}; therefore, as a first-order approximation, they are not included in the surrounding winds.


\subsection{Non-Thermal Proton Density}

Neutrino emission from the jets is taken to originate from  proton--proton interaction between a distribution of hot (fast) protons and cold (bulk flow) protons \cite{Romero2003,Reynoso2008,Reynoso2009,Reynoso2019,Kelner2006,Lipari2007}. Some of the bulk protons are accelerated at shock fronts according to the first-order Fermi acceleration mechanism, with a time scale of \cite{Begelman1980,Rieger2006,Rieger2019}

\begin{equation}
t^{-1}_{\mathrm{acc}} \simeq \eta \frac{c e B}{E_{p}},
\end{equation}
where {\emph{{B}} 
is the magnetic field}, { $E_p$ is the proton energy,} $e$ the proton charge, and $c$ the speed of light. $\eta=0.1$ represents an acceleration efficiency parameter, assuming efficient acceleration in moderately relativistic shocks in the vicinity of the jet base \cite{Begelman1980}.

As an approximation,  high-energy electron distribution is deferred to future work. Focusing on hadrons, we adopt a power-law distribution for the relativistic protons {as a function of their energy \emph{E}} of the form $N_p = K N_{p(0)} E^{-\alpha}$ \cite{Hughes1991},  {\emph{K} being a scaling constant connecting fast hot and thermal proton densities, $N_p$ being the bulk proton density at a given jet point and $N_{p(0)}$ the density at a reference jet point}, with either $\alpha \approx 2 $ \cite{Reynoso2008}, or a variable $\alpha$ \cite{Reynoso2019}, where $\alpha$ is the proton spectral index in the local jet cell matter frame. Alternatively, a transport equation could be used to find the distribution \cite{Reynoso2008}.

As a further approximation, the aforementioned hot proton distribution is taken to be isotropic in the jet frame, assuming that,  {at each jet point,}  $l_{sc} < l_{r}$, where $l_{sc}$ is the scattering length  and $l_{r}$ the radiative length, a hypothesis backed by the nature of diffuse shock acceleration \cite{Derishev06}.

\subsection{A Note on Jet Frame Anisotropy}
\label{A NOTE ON ASSUMED JET-FRAME ISOTROPY EMISSION.}


For protons accelerated at diffuse shocks, the above assumption of isotropy is justified by the need to preserve, after every bounce, at least some proton energy \cite{Rieger2019}. Consequently,  scattering length $l_{sc}$ is less than  radiative length $l_{r}$. Otherwise, the proton would not have any energy left after the bounce, negating the acceleration process.

According to the work in \cite{Derishev06},  assumed anisotropy of  hot proton distribution can be reflected to  neutrino distribution. In the jet system, emission would then be projected off axis, even reinforced, under certain orientations, in the lab frame.


\section{Neutrino Emission Calculations}
\label{neutrinos}

\subsection{Proton Energy Loss }

\textls[-15]{Following the works in \cite{Kelner2006,Reynoso2009,Reynoso2019}, certain energy loss mechanisms are included. This presentation is performed in a cell with the properties of ($u_{x}, u_{y}, u_{z}, b_{x}, b_{y}, b_{z}, n, \phi_{1}, \phi_{2}, \alpha$)~=~($-$0.3780c, 0.4480c, 0.0124c, 10$^{5}$ G, 10$^{6}$ G, 10$^{5}$ G, 2.1 $\times$ 10$^{11}$ cm$^{-3}$, 1.047 rad, 5.00~$\times$ 10$^{-7}$ rad, 2.0). In the latter, \emph{u} stands for velocity and \emph{b} for magnetic field along directions \emph{x}, \emph{y}, or \emph{z}. Bulk flow proton density is denoted by \emph{n}, and $\phi1 \simeq \phi$ ($\phi_{2}$~=~0) is the complementary to $\theta$, the angle to the line of sight. Last, $\alpha$ is the high-energy proton distribution spectral index. As an exception, the pion injection function presentation uses a different velocity vector of (0.2, 0.8, 0.1)c. Nevertheless, in the model runs, these are potentially performed in every cell.}



We consider  cut-off E for protons E $\le$ 10$^{6}$ GeV.
For the adiabatic expansion time scale, we have \cite{Reynoso2009}
\begin{equation}
t_{\mathrm{adb}}^{-1}=\frac{2}{3} \frac{ u_{ b(\mathrm{adb})} }{ z_{j} }
\end{equation}
where z$_{j}$ = 10$^{11}$ cm is the characteristic lateral size scale of the jet. For this simple calculation, $u_{{b}(\mathrm{adb})}$ is preset to 0.8c.

\begin{figure}[H]
\includegraphics[width=10.5 cm]{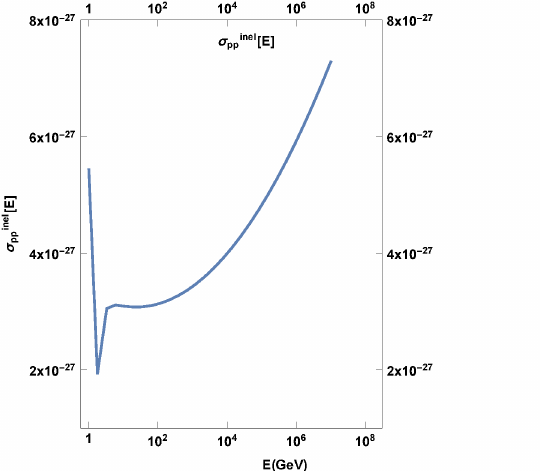}
\caption{Inelastic proton--proton collision standard plotted with energy. {It demonstrates  rather small variation (linear vertical scale) of its value over a large energy range (logarithmic horizontal scale), covering and exceeding the energy span required for the calculations that follow later in this paper. } }
\label{sigma_inel_pp}
\end{figure}

For the \emph{{p}}--\emph{{p}} collision loss mechanism, we have
\begin{equation}
t_{pp}^{-1}=n c \sigma_{\mathrm{inel_{pp}}}(E_{p}) K_{pp}
\label{tpp1}
\end{equation}
where \emph{n} is the bulk flow proton number density, $K_{pp}$~=~0.5 \cite{Reynoso2009} and $\sigma^{(\mathrm{inel_{pp}})}_{pp}$ is the inelastic \emph{{p}}--\emph{{p}} collision cross section \cite{Reynoso2009}
\begin{equation}
\sigma^{(\mathrm{inel_{pp}})}_{pp}=(34.3+1.88L+0.25L^{2}) \times [1-(\frac{E_{th}}{E_{p}})^{4}]^{2} \times 10^{-27} \mathrm{cm^{2}}
\end{equation}
where ${E_{th}}$~=~1.2 GeV and {\emph{L}~=~ln(\emph{E}$_{p}$/1000~GeV))
(see in \cite{Reynoso2009,Kelner2006}). Equation \eqref{tpp1} is justified if we consider a small cube of matter of number density \emph{n}, moving at speed (near) \emph{c} and having a surface A perpendicular to its direction of motion. Then, \emph{n} $\times$ \emph{c} has the dimensions of cm$^{-2}$ $\times$ s$^{-1}$. This is then multiplied by $\sigma^{(\mathrm{inel_{pp}})}_{pp}$, resulting in the inverse time scale for the aforementioned \emph{p}--\emph{p} collision.}
In Figure \ref{sigma_inel_pp}, $\sigma^{(\mathrm{inel})}_{pp}$ is plotted.

For the pion decay time $t_{\pi}$ and the characteristic pion decay timescale $t_{\pi 0}$, we have {the following equations:}
\begin{equation}
t_{\pi 0}=2.6 \times 10^{-8} \mathrm{s}
\end{equation}
and
\begin{equation}
t_{\pi}=t_{\pi 0} \Gamma _{\pi} +t_{\mathrm{esc}}
\end{equation}
{where $\Gamma_{\pi}$ is the pion Lorentz factor,}

which, in practice, takes the form {(\emph{m}$_{\pi}$ is the pion mass and $E_{\pi}$ the pion energy)}
\begin{equation}
t_{\pi}=t_{\pi 0} (\frac{E_{\pi}}{m_{\pi} c^{2}}) +t_{\mathrm{esc}}
\end{equation}
where  light escape time $t_{\mathrm{esc}}$ strongly affects the final result.

The synchrotron loss time scale is defined by {\cite{Reynoso2009} }
\begin{equation}
t_{\mathrm{sync}}^{-1}=\frac{4}{3} (\frac{m_{e}}{m_{p}})^{3} \frac{1}{8 \pi c m_{e}} \sigma_{T} B^{2} \frac{E_{p}}{m_{p} c^{2}}
\end{equation}
{{$m$}$_{e}$ is the electron mass and {$m$}$_{p}$ the proton mass. $\sigma_{T}$ = $\frac{8 \pi}{3} (\frac{e^{2}}{m_{e} c^{2}})^{2}$ = 6.65 $\times$ 10$^{-25}$ cm$^{2}$ is the Thompson cross section, \emph{e} is the electron charge and \emph{B} is the local magnetic field.} The form of the {latter} term {$\frac{E_{p}}{m_{p} c^{2}}$, which is equal to the proton Lorentz factor,} {which is essentially }$\Gamma_{p}$, facilitates energy-dependent calculations later. In total,
\begin{equation}
t_{\mathrm{loss}}^{-1}=t_{\mathrm{sync}}^{-1}+t_{\mathrm{adb}}^{-1}+t_{pp}^{-1}
\end{equation}

\begin{figure}[H]
\includegraphics[width=10.5 cm]{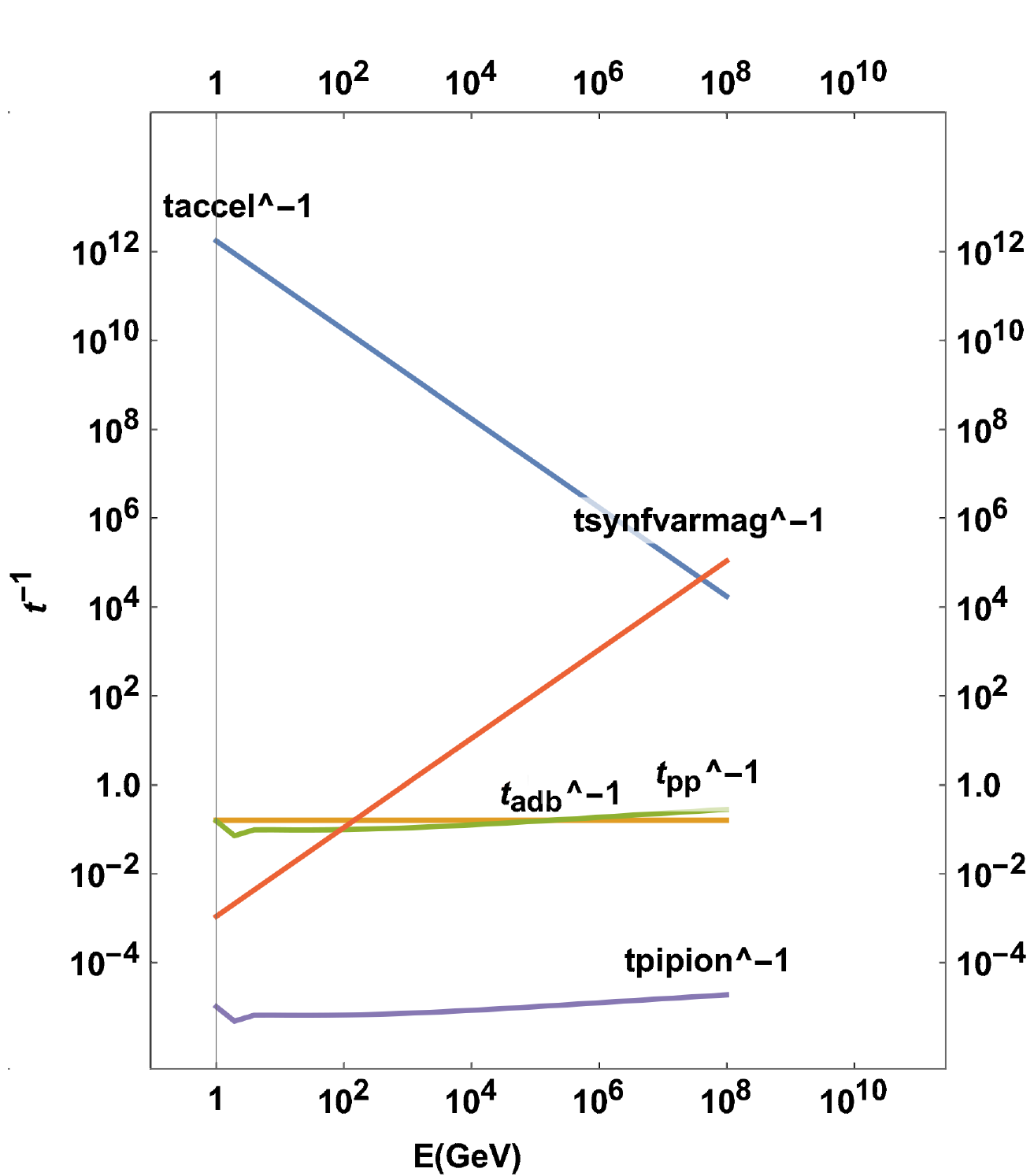}
\caption{High-energy proton distribution loss time scales, for various processes in the jets, plotted with energy in { GeV. t$_{\mathrm{accel}}$ is the proton acceleration time scale at shocks. t$_{\mathrm{synfvarmag}}$ stands for the synchrotron mechanism loss time scale, using a magnetic field that varies from point to point within the jet. t$_{\mathrm{adb}}$ is the adiabatic loss time scale, t$_{\mathrm{pp}}$ is the (hot--cold) proton--proton collision timescale.} t$_{\mathrm{pipion}}$ stands for the pion decay timescale t$_{\pi}$.}
\label{tloss_many}
\end{figure}

{In Figure \ref{tloss_many}, the various energy-loss mechanism time scales  are presented.}

\subsection{Model for the Interaction of Thermal and Non-Thermal Protons in the Jet}

Hot--cold proton interaction results to a distribution of high-energy pions, which then decay, allowing for the creation of energetic neutrinos. We have \cite{KS18p,SK17,SK15,Campion2020}
\begin{equation}
pp \rightarrow pp \pi^{0}+\pi_{0} \, ,
\end{equation}
for neutral pions $\pi^{0}$, and
\begin{equation}
pp \rightarrow pn \pi^{+} {+\pi_{+} }\, , \qquad pp \rightarrow pn \pi^{-}+ {\pi_{-} } {\pi^{+}+\pi^{+} }\, ,
\end{equation}
for $\pi^{\pm}$.

$\pi^0$ decay to gamma rays, while $\pi^{\pm}$ mostly decay to an antimuon or muon and a
muonic neutrino or antineutrino (prompt neutrinos) \cite{KS18p,Campion2020}.
\begin{equation}
\pi^{+} \rightarrow	\mu^{+} + \nu_{\mu} \, ,  \qquad
\pi^{-} \rightarrow \mu^{-} + \widetilde{\nu}_{\mu} \, .
\label{muon-produc}
\end{equation}

As an approximation, we neglect both neutrino production through secondary channels and delayed neutrinos.

\textls[-8]{For each successive particle population in the above cascades,  the transport \mbox{equation \cite{Duderstadt1979}} can be solved.}

{The transport equation for nonstochastic phenomena and for time-independent transport (transport time much less than the time step of the dynamic simulation) takes the following simplified form:
\begin{equation}
\frac{\partial N}{\partial E} + \frac{N}{t_{\mathrm{loss}}} = Q(E,\vec{r})
\end{equation}
where $t_{\mathrm{loss}}$ is the decay timescale for the particle in question, $\vec{r}$ is the location vector in space, \emph{N} is the particle density of the produced particle population, and \emph{Q} is its injection function. \emph{Q} is calculated from the previous population up the cascade. For example, \emph{N} may represent protons and \emph{Q}, {(which includes proton acceleration effects),} is expressed as a function of  the local thermal proton distribution. In turn, the \emph{N} of pions can be obtained using their \emph{Q}, which in turn is a function of the hot proton \emph{N}, and so on along the cascade.}

Nevertheless, a power-law distribution is assumed for protons, skipping having to solve the first transport equation in the cascade. From protons to pions, then to muons and neutrinos, each generation of particles leads to the next one. The authors of \cite{Kelner2006} calculated the properties of resulting particle distributions over a large energy range, performing Monte Carlo calculations with the results of particle physics.

\subsubsection*{ PLUTO Code}
PLUTO \cite{Mignone2007} is an open-source, 2D/3D modular hydrocode, a finite-volume/-difference shock-capturing program, meant to integrate a set of (time-dependent) conservation laws. Initial and boundary conditions are conveniently assigned through an equivalent set of primitive variables. The relevant systems of equations may include hydrodynamics (HD), magnetohydrodynamics (MHD), and their special-relativistic counterparts, RHD and RMHD, respectively, in either two or three spatial dimensions. The solution of conservation laws is carried out through discretization on a structured mesh, a logically rectangular grid surrounded by a boundary with additional ghost cells in order to implement boundary conditions. The grid may either be static or adaptive, and various coordinate systems are available. The programme may run efficiently in parallel on various platforms.

In previous works \cite{SK15,SK14,SK11}, the hadronic jet was
modelled using the PLUTO code. PLUTO results   were then processed in order to calculate the emissivity of $\gamma$ rays and neutrinos using various approximations. {As far as neutrinos are concerned,  emission  was calculated at only a handful of points along the jet, thus not taking advantage of the detail offered by a numerical jet simulation.} In this paper, the emissivity of neutrinos is separately calculated  at each computational cell using the angle (los,u) formed between  LOS and  local velocity. {This calculation is huge compared to the previous one, but the benefit is that a result is obtained at each point. New code NEMISS \cite{NEMISS} (not available before) is employed here, which performs the calculation on data produced by PLUTO. }

{Furthermore, a time-delay-capable line-of-sight code, RLOS2 \cite{RLOS}, is employed here, which was not available in the aforementioned previous works (much simpler LOS code~\cite{los_code} used back then was also written by this author). RLOS2 reads the combined results of PLUTO and of NEMISS, and produces synthetic neutrino images of the model system using either a focused beam geometry or parallel rays. }

{The improved relativistic transformation of the hot proton distribution by \cite{TR11} is now employed as opposed to \cite{PS2001} in the previous works. }

{The PLUTO jet is now a twin in 3D space, using new files for setting up PLUTO for our problem \cite{twin_mq_jets}, adding to the realism of the system. The twin jet model system {employed here} is an evolution of the single jet system, also set up by this author in PLUTO \cite{mq_jet}. The magnetic field is now adjusted for equipartition, and the model parameters are generally more refined   compared to those in earlier works. }


\subsection{Lorentz Transform of High E Proton Distribution}

For the calculation of the fast proton distribution, the relevant directional equation {(direction is defined by the angle $\theta$ between  velocity and  line of sight)} is found in \cite{TR11,PS2001}. 
The latest variant originates from \cite{TR11} (TR), used here, minus a geometry factor that we absorb into the normalization factor
\begin{equation}
n(E,\theta)=\frac{\Gamma^{- \alpha -1} E^{- \alpha} (1- \beta \mathrm{cos}(\theta) \sqrt{1- \frac{m^{2} c^{4}}{E^{2}} } )^{- \alpha -1} }{[\mathrm{sin^{2}}(\theta) + \Gamma^{2}(\mathrm{cos}(\theta) - \frac{\beta}{\sqrt{1- \frac{m^{2} c^{4} }{E^{2}}}})^{2}]^{\frac{1}{2}}  }
\label{tr13}
\end{equation}
where $\Gamma$ is the Lorentz factor of the particles in a tiny volume. The particle mass is m and $\beta$~=~\emph{u}/\emph{c} is the particle speed in units of the speed of light \emph{c}. \emph{E} is the particle energy and $\alpha$ is the spectral index of the distribution.

On the other hand, {the authors of} 
\cite{PS2001} (PS) say
\begin{equation}
n(E,\theta)=\frac{\Gamma^{- \alpha +1} E^{- \alpha} (1- \beta \mathrm{cos}(\theta) \sqrt{1- \frac{m^{2} c^{4}}{E^{2}} } )^{- \alpha} }{[\mathrm{sin^{2}}(\theta) + \Gamma^{2}(\mathrm{cos}(\theta) - \frac{\beta}{\sqrt{1- \frac{m^{2} c^{4} }{E^{2}}}})^{2}]^{\frac{1}{2}}  }
\label{ps01}
\end{equation}

A simpler variant is \cite{SK14}
\begin{equation}
n(E, \theta)=\Gamma(E-\beta \sqrt{E^{2}- m^{2} c^{4} \mathrm{cos}(\theta) })
\label{simpler_variant}
\end{equation}

\begin{figure}[H]
\includegraphics[width=10.5 cm]{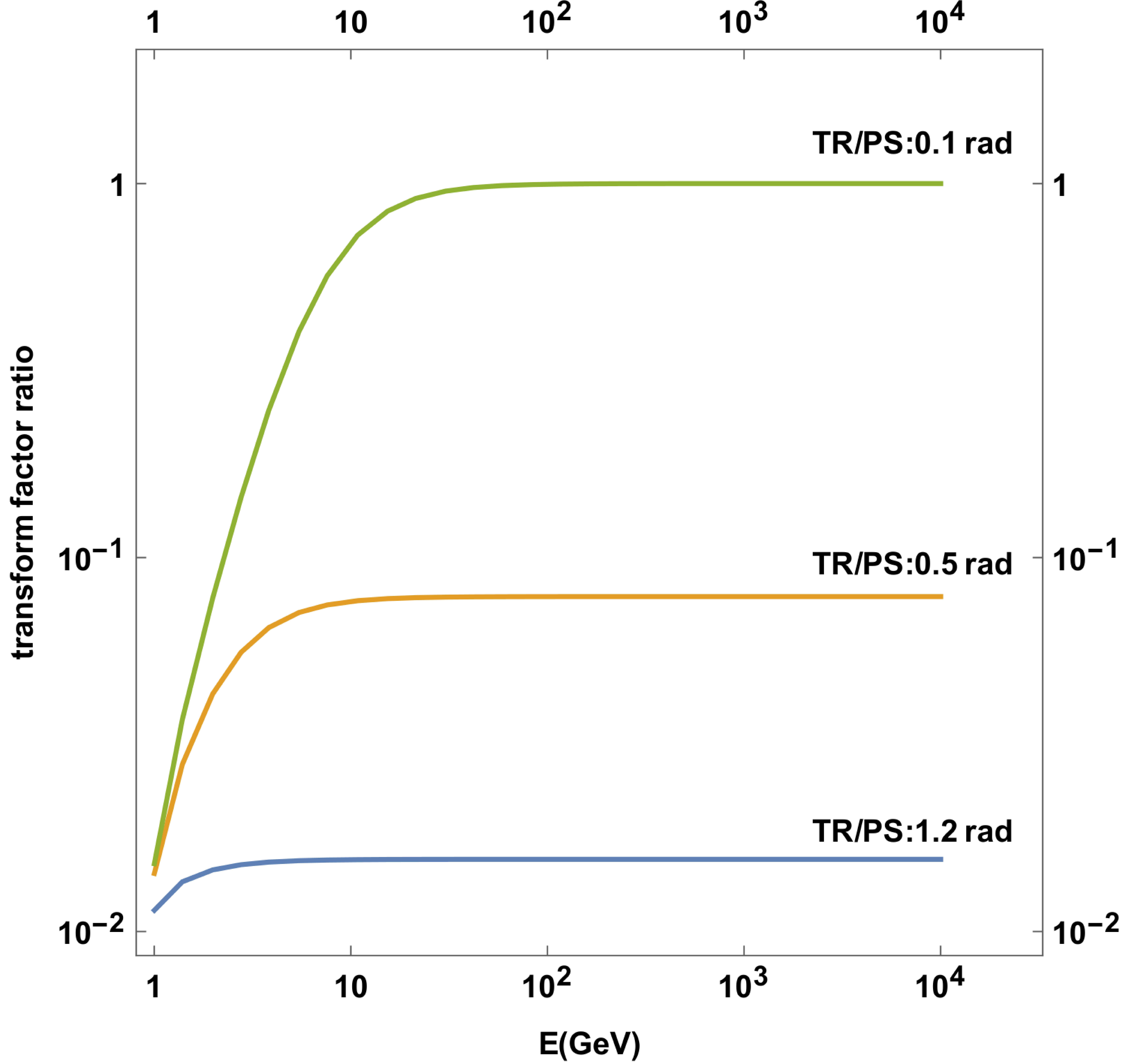}
\caption{Ratio of high-energy proton distribution density transformation {as calculated by the formulae of TR and PS, respectively,} for three different angles. {{TR stands for} 
\cite{TR11}, PS for \cite{PS2001}.} A reproduction, for verification, of a figure from in \cite{TR11}.  }
\label{tr13ps01angles}
\end{figure}

\begin{figure}[H]
\includegraphics[width=10.5 cm]{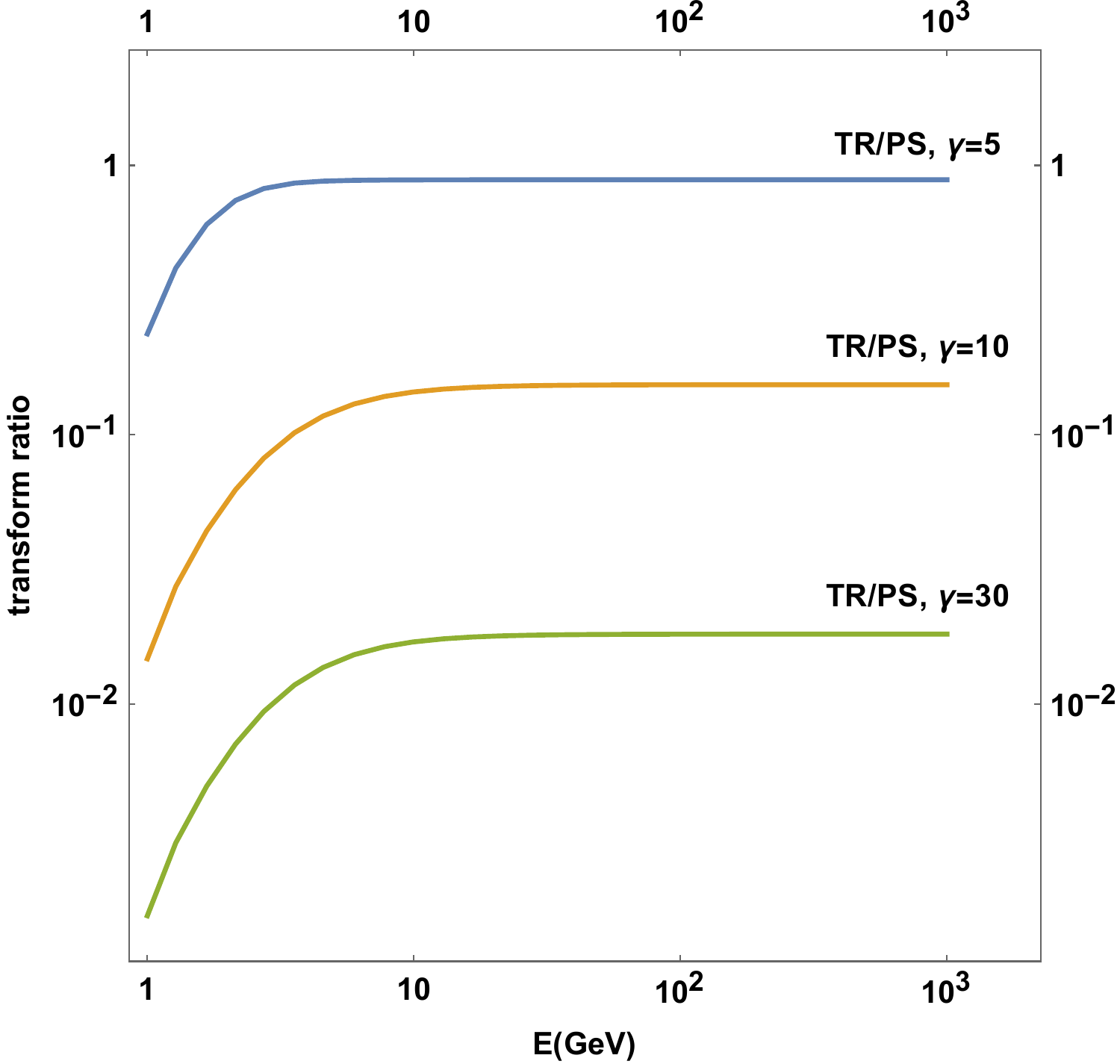}
\caption{Ratio {of the results of TR and PS} {of TR/PS} non-thermal proton distribution density transformation factors for three different Lorentz factors. {{TR stands for} \cite{TR11}, PS for \cite{PS2001}.} A reproduction, for verification, of a figure from in \cite{TR11}.}
\label{tr13ps01lorentz}
\end{figure}

Equations  \eqref{tr13} and \eqref{ps01} are compared in Figures \ref{tr13ps01angles} and \ref{tr13ps01lorentz}.

%
%

\subsection{Pion Injection Function and Pion Energy Distribution}


For each fast--slow proton interaction, a spectrum of possible pion energies exists, given by  function $F_{\pi}$ \cite{Kelner2006,Reynoso2009,Reynoso2019}.

\nointerlineskip
\begin{eqnarray}
F_{\pi}^{(pp)}\left( x,\frac{E}{x} \right) = 4 \alpha B_{\pi} x^{\alpha-1}
\left( \frac{1-x^{\alpha}}{1+r x^{\alpha}(1-x^{\alpha})} \right)^{4}
\left( \frac{1}{1-x^{\alpha}} + \frac{r(1-2x^{\alpha})}{1+rx^{\alpha}(1-x^{\alpha})}
\right) \left( 1- \frac{m_{\pi} c^{2}}{x E_{p}}  \right)^{\frac{1}{2}}
\label{Ffunction}
\end{eqnarray}
\noindent where
$x=E/E_{p}$.

\begin{figure}[H]
\includegraphics[width=10.5 cm]{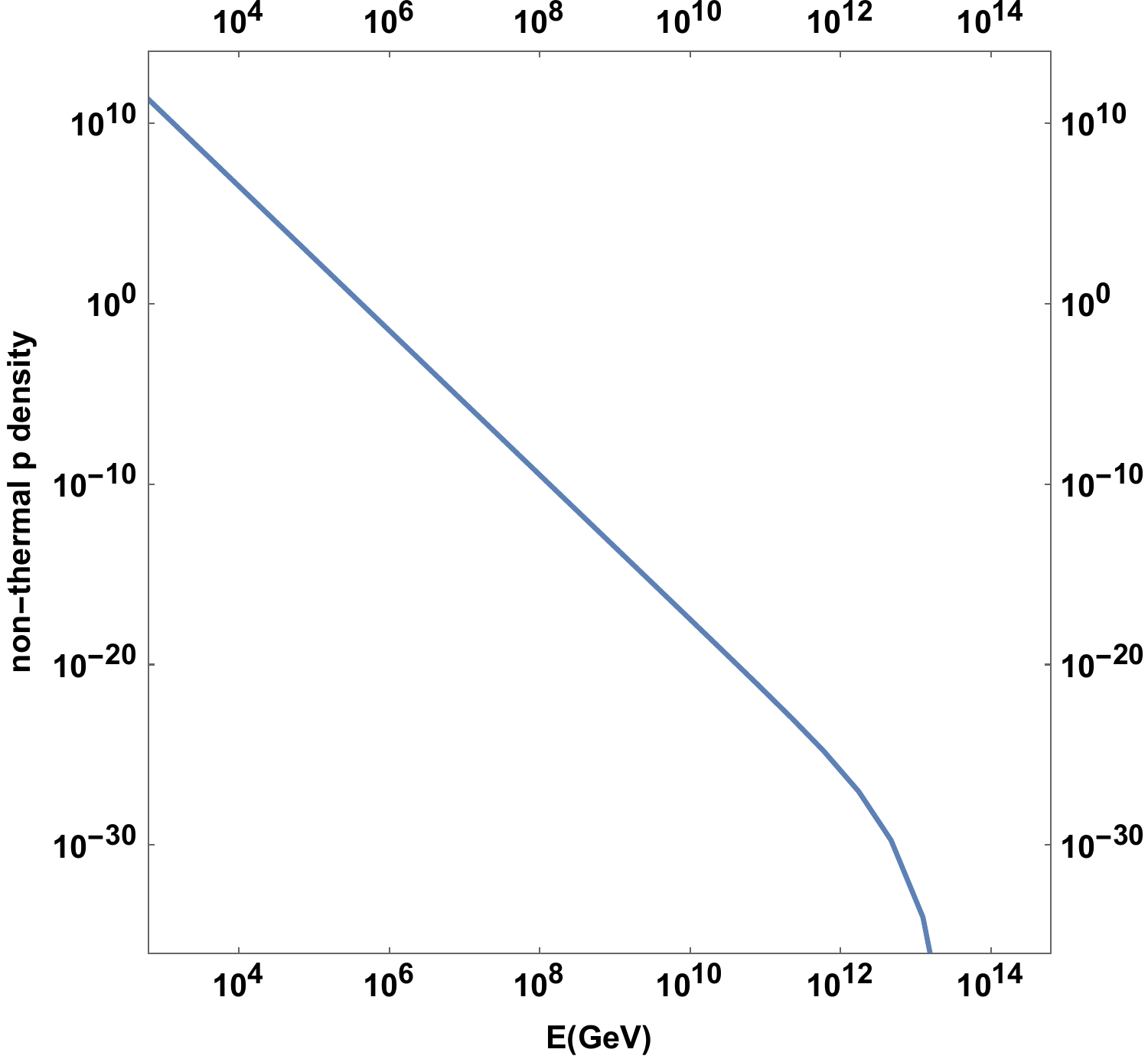}
\caption{\textls[-15]{Density of non-thermal protons in the jet using a high-energy cut-off feature plotted \mbox{with~energy}.}}
\label{fast-p-density}
\end{figure}

\begin{figure}[H]
\includegraphics[width=10.5 cm]{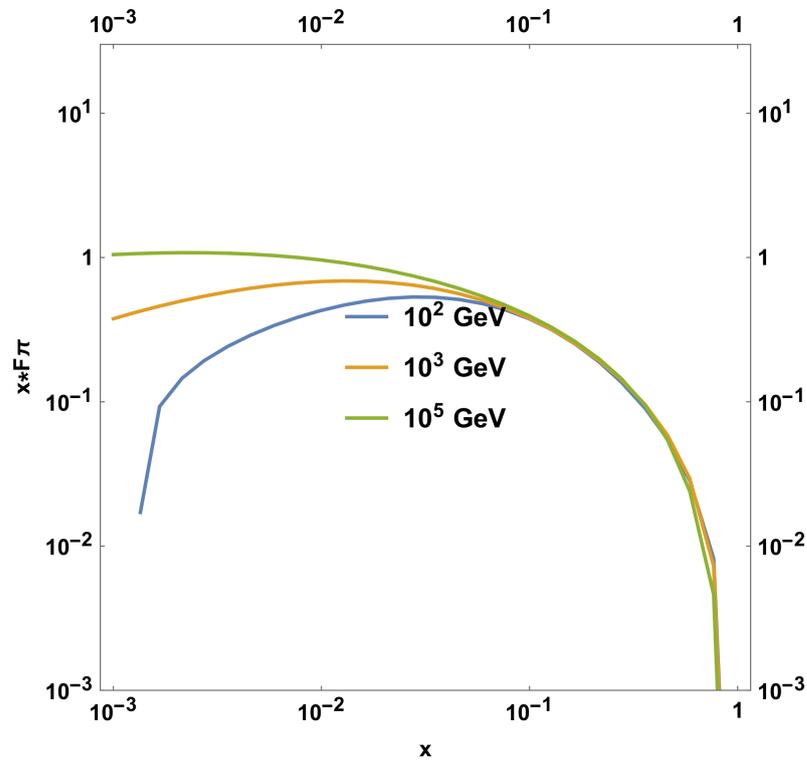}
\caption{{ \emph{F}$_{\pi}^{(pp)}\left(x,\frac{E}{x} \right)$ function,} Equation \eqref{Ffunction}, { F function}, corresponding to the pion spectrum emerging from a single (hot--cold) proton collision,  multiplied by the \emph{x}~=~$\frac{E_{\pi}}{E{p}}$ fraction. Calculation  performed at three different energies for the non-thermal proton.}
\label{Fpp_plot}
\end{figure}

Figure \ref{fast-p-density} shows  p-law fast proton density.  Figure \ref{Fpp_plot} $xF_{\pi}$ is plotted with the fraction
$x$ for different fast proton energies.

Pion injection function $Q_{\pi}^{(pp)}$ comprises pion contributions at each pion energy to that pion energy from spectrum $F_{\pi}^{(pp)}\left( x,\frac{E}{x} \right)$ of all potential \emph{p}--\emph{p} interactions.

\begin{eqnarray}
Q_{\pi}^{(pp)}(E,\vec{{r}} {z}) = n(\vec{{r}} {z}) c \int \limits_{\frac{E}{E_{p}^{(max)}}}^{1} \frac{dx}{x}
\left(  \frac{E}{x},\vec{{r}}  {z} \right) F_{\pi}^{(pp)} \left( x,\frac{E}{x} \right)
\sigma^{(inel)}_{pp} \left( \frac{E}{x} \right)  \, ,
\label{Qpp}
\end{eqnarray}
$x$ is the fraction of the pion energy to proton energy, and $n(\vec{{r}} {z})$ is the jet flow proton density.

\vspace{-6pt}

\begin{figure}[H]
\includegraphics[width=10.5 cm]{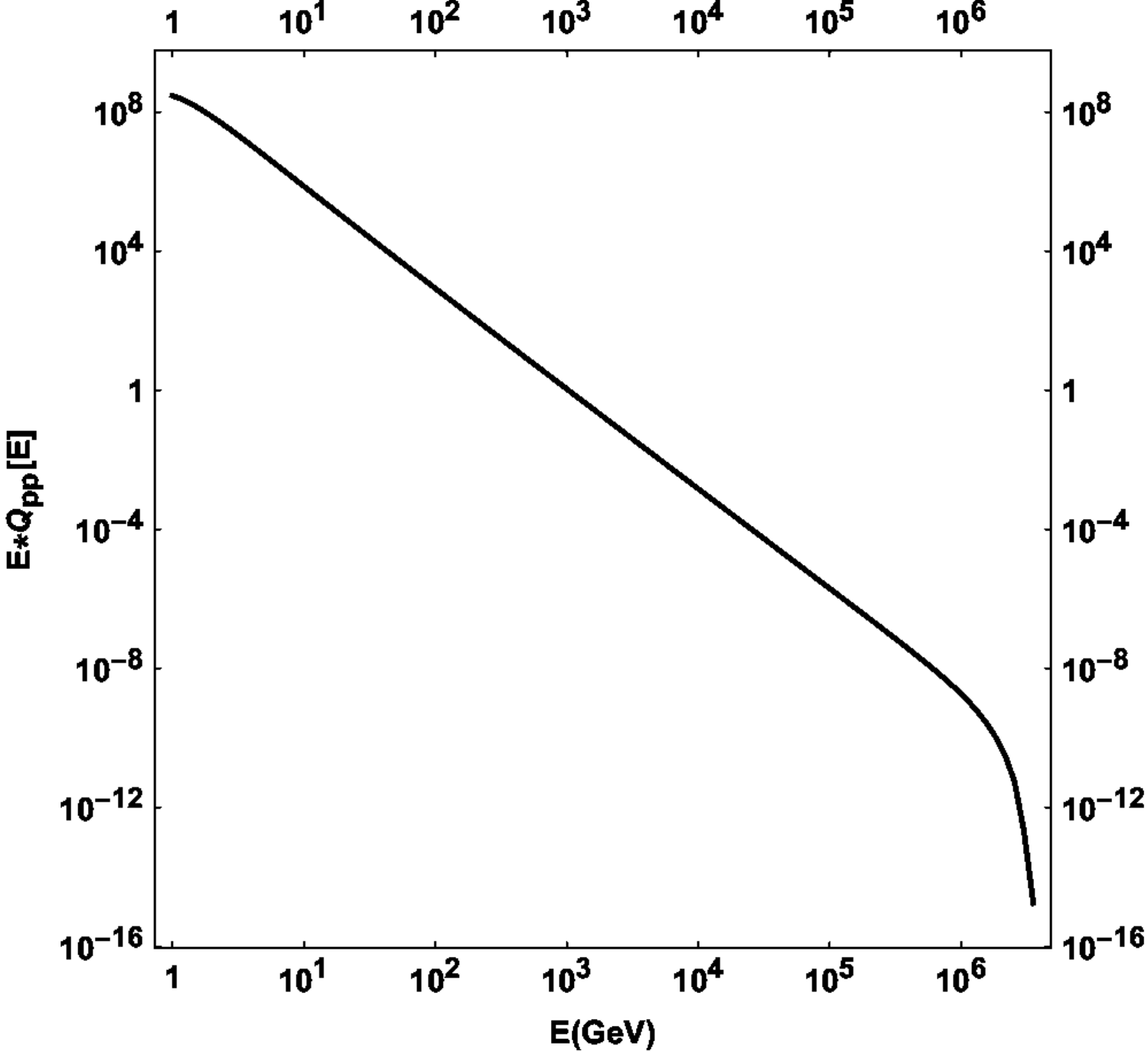}
\caption{Pion injection function Q, weighted by pion energy, measured in non-normalized units, describing the combined spectrum from a multitude of (hot--cold) \emph{p}--\emph{p} collisions. We can see contributions rapidly declining as particle energy increases. {As an exception, this figure uses a velocity vector of (0.2, 0.8, 0.1)c. } }
\label{qpp_plot}
\end{figure}

\clearpage
Figure \ref{qpp_plot} plots $Q_{\pi}^{(pp)}$ versus pion
energy $E_\pi$.

In order to obtain  pion distribution, we solve the following transport equation:
\begin{equation}
\frac{\partial N_{\pi}}{\partial E} + \frac{N_{\pi}}{t_{\pi}} = Q_{\pi}^{(pp)}(E,\vec{{r}} {z})
\end{equation}
where \emph{N}$_{\pi}(E,\vec{{r}})$ denotes the pion energy distribution. We proceed
\begin{eqnarray}
N_{\pi}(E) = \frac{1}{|b_{\pi}(E)|} \int \limits_{E}^{E^{(max)}} dE' Q_{\pi}^{(pp)}(E')
\exp {[-\tau_{\pi}(E,E')]} \, ,
\end{eqnarray}
where
\begin{eqnarray}
\tau_{\pi}(E',E)=\int \limits_{E'}^{E} \frac{dE'' t_{\pi}^{-1}(E)}{|b_{\pi}(E'')|} \, .
\end{eqnarray}

\begin{figure}[H]
\includegraphics[width=10.5 cm]{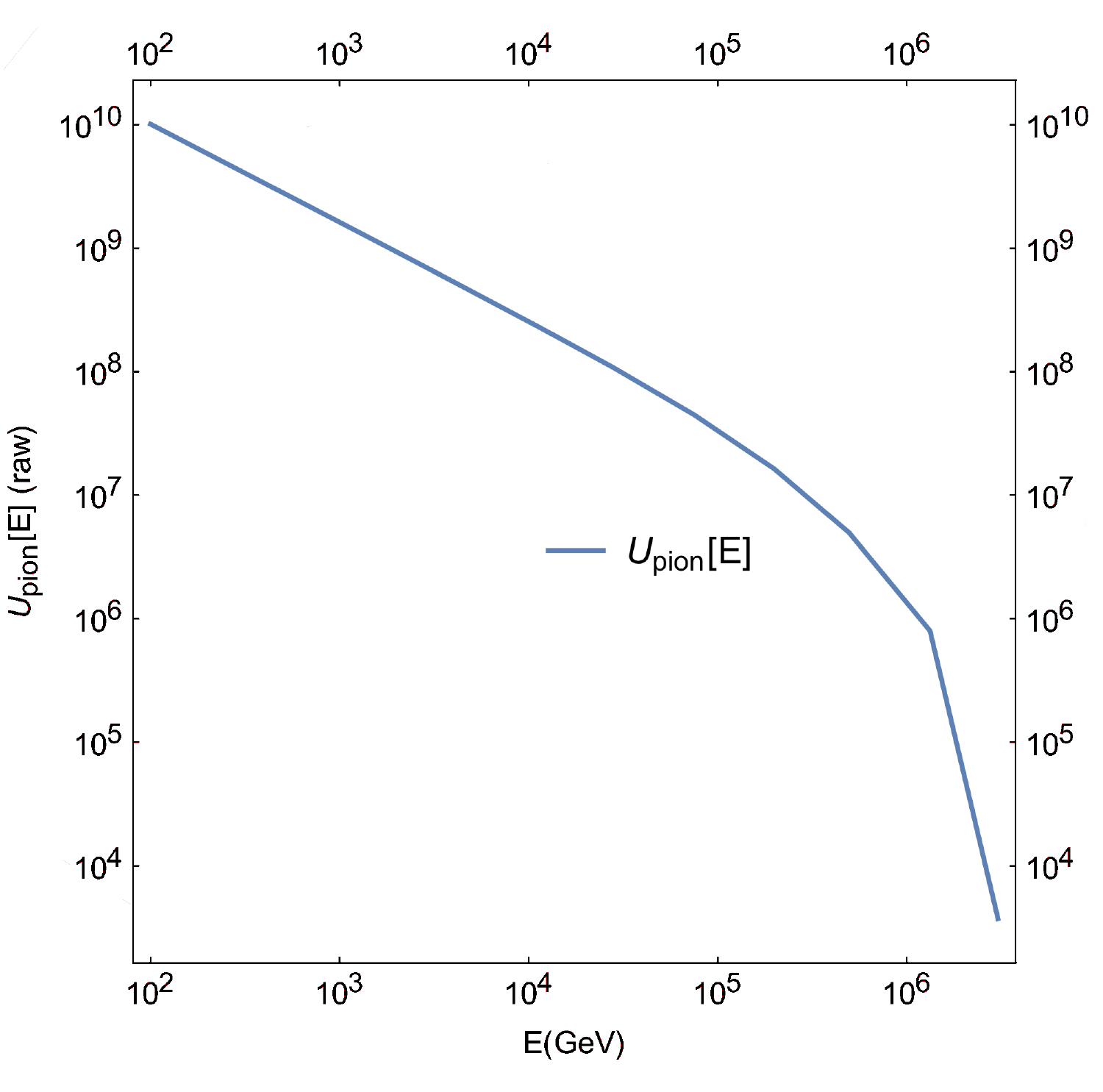}
\caption{Pion energy distribution plotted in non-normalized units versus energy. {In the software, the above distribution is represented by  function U.} }
\label{upp_plot}
\end{figure}

The quantity $\tau_{\pi}(E',E)$ is the pion optical depth and $b_{\pi (E)}=-E(t_{\mathrm{sync}}^{-1}+t_{\mathrm{adb}}^{-1}+t^{-1}_{\pi p}+t^{-1}_{\pi \gamma}) $ is the energy loss rate of the pion. As an approximation, the last term in the latter expression is omitted. {Figure \ref{upp_plot} plots  NEMISS software function U (U$_{\mathrm{analytical}}$), representing \emph{N}$_{\pi}$, with pion energy.}

The above are performed for each computational cell, where the quantities for radiative purposes are considered locally constant.
A cell is macroscopically large inasmuch as only the deterministic portion of the transport equation is
employed, in turn rendering it deterministic. Again, we take the
characteristic scale (mean free path) of the radiative interactions to be smaller than
the cell size, leading to the containment of particle interactions within a given
hydrocode cell. Furthermore, the time scale for the radiative interactions is taken to be
smaller enough than the hydrocode's time step, so that the radiative interactions belong to a
single time step each time.


\subsection{Neutrino Emissivity}

The emissivity of prompt neutrinos \cite{Kelner2006,Lipari2007,Reynoso2008,Reynoso2009} is
\vspace{-3pt}
\begin{eqnarray}
Q_{\pi \rightarrow \nu}(E) = \int \limits_{E}^{E_{max}} dE_{\pi} t^{-1}_{\pi}
(E_{\pi}) N_{\pi}(E_{\pi}) \frac{\Theta (1-r_{\pi}-x)} {E_{\pi}(1-r_{\pi})}  \, ,
\label{Neut-Emiss}
\end{eqnarray}
{where \emph{E} is neutrino energy, $r_{\pi}=(m_{\mu}/m_{\pi})^{2}$,} $x=E/E_{\pi}$, and $t_{\pi}$ is the pion decay timescale. $\Theta$($\chi$)
is the theta function ~\cite{Reynoso2009,SK15}.
Neutrino emissivity is calculated for each individual cell using the cell's own angle to the LOS crossing that cell. The imaging process may  incorporate either parallel LOSs or a focused beam, where each LOS follows a slightly different path to a common focal point \cite{RLOS}. A synthetic image of the model system is \mbox{thus produced}.

\section{Results and Discussion}
\label{results}



{Using the formalism   presented in this paper so far, the neutrino emission at each computational cell of the model  is calculated. This method is heavier from a computational point of view, but allows for obtaining a separate neutrino emission from each spatiotemporal point of the twin jet model. Thus, we aim for the result
\begin{equation}
I_{\nu}=I_{\nu}(\vec{r},t)
\end{equation}
where intensity \emph{I} is calculated at the 3D computational cell at $\vec{{r}}$, represented by the x, y, and z coordinates of the cell. Time t is obtained from the time tag of the PLUTO data dump where the cell belongs. Thus, the above equation is globally applied  to all selected PLUTO data (the user may select beginning and end times for the global calculation). We now proceed to describe the setup of the simulation.  }

The jet base is situated near the centre of a Cartesian grid.
A continuous model jet representing a microquasar system is injected at a u$_{\mathrm{jet}}$ = 0.865c {(a Lorentz factor of 2, which lies between higher microquasar Lorentz factors used in the literature, such as $\Gamma$~=~5 in \cite{Romero2003}, and $\Gamma$~=~5/3, corresponding to u = 0.8c, a characteristic value for the jets in GRS1915~+~105) } is studied with the RMHD setup of the PLUTO hydrocode, at a uniform grid resolution of 60 $\times$ 100 $\times$ 50. Grid size is (120 $\times$10$^{10}$\, cm) $\times$ (200 $\times$10$^{10}$\, cm)  $\times$ (100 $\times$10$^{10}$\, cm); therefore, cell length is 2 $\times$10$^{10}$\, cm. {The grid size is such that it focuses on the area of the inner jet, where $\gamma$ ray and neutrino production is expected. This minimal size of the cell means that a starting radius of the jet of a few times 10$^{10}$ cm is necessarily implied, as a few cells' diameter of the nozzle is used. This compromise is imposed by the nature of the employed simulation, which utilises a homogeneous grid. In future work, a non-homogeneous grid may allow for better focusing on more realistically resolving the jet input nozzle.}

\begin{figure}[H]
\includegraphics[width=10.5 cm]{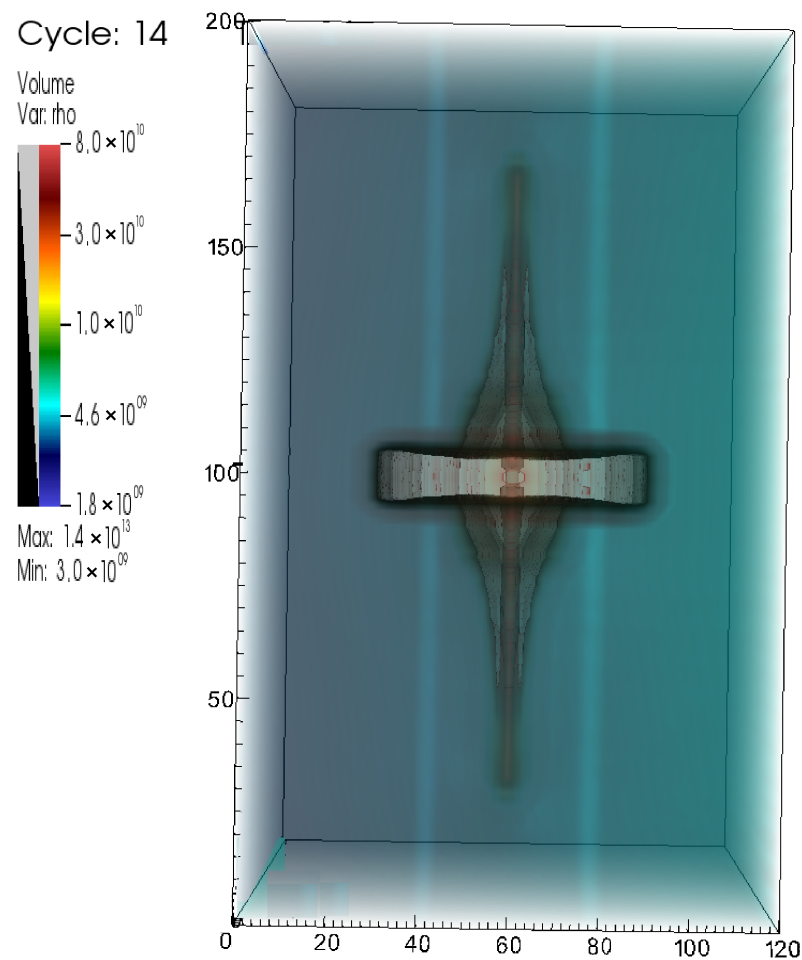}
\caption{{Three-dimensional} 
side view of  twin model jet system. Snapshot 14 of  u = 0.866c hydrocode run corresponding to a model time of t = 350 s (14 $ \times $ 25), depicting the density in a logarithmic plot. Both jet fronts are advancing towards the ends of the grid, traversing the surrounding stellar wind after crossing the simplified accretion disk wind construct. Image produced with VisIt.}
\label{visit_shot1}
\end{figure}

In all of the model runs, the same initial jet density of 10$^{10}$ protons/cm$^{3}$ was used {(a typical value for the inner microquasar jet, also compatible with the energetics of the jet and its kinetic luminosity)}, 2000 times less than the maximal surrounding gas density {(i.e., a light jet is assumed, which is a possibility that supports a rich jet--wind interaction environment, but is also more demanding from a computational point of view)}. Winds comprise an accretion disk wind construct and a stellar wind that  falls off away from the companion star, located off-grid at (4~$\times$~10$^{12}$\, cm, 1~$\times$~10$^{12}$\, cm, 4~$\times$~10$^{12}$\, cm), while the jet is threaded by a strong confining toroidal magnetic field of \emph{B}~=~10$^{4}$ G, assuming equipartition between kinetic and magnetic energy density (see Appendix \ref{Equipartition_calculation} for the calculation of the latter equipartition value for B, in relation to the jet kinetic luminosity). A guide for  inner system winds and their densities was SS433 \cite{Fabrika2004}. Simulations were run until t = 842 s, saving a data snapshot every 25 (simulation) s. A three-dimensional  snapshot of density is shown in Figure \ref{visit_shot1}, where we can see the magnetically collimated jet pair advancing through surrounding winds.

%

{The above figures show a narrow jet barely expanding into its surrounding winds. This small half-angle is then rather counterintuitively expected to result in a faster decline of neutrino emission with energy, as discussed in the discussion section of \cite{Reynoso2009}. }

A number of empty user parameters were employed in order to house particle emission results later. Then, the above PLUTO run was copied into many directories. In each, the NEMISS programme \cite{NEMISS} was run, which calculates neutrino emissions for a  specific imaging geometry and setup. { This programme is able to read 4D spatiotemporal data output from PLUTO into a 5D array, which also includes particle energy as a fifth dimension. Then, NEMISS calculates the neutrino emission at each point of the 5D data array.} Results were overwritten into suitably prepared data files of the originally empty user parameters of the hydrocode. { Thus, NEMISS processes PLUTO output to include a neutrino emission spectrum at each spatiotemporal data point.}

{PLUTO data processed by NEMISS are then ready to be read by  relativistic time-delay LOS imaging programme RLOS2 \cite{RLOS}, which produces synthetic neutrino images of the system. Over a string of particle energies, the intensity sum of the whole synthetic image of the jets is calculated for each energy. Thus, the plot of jet neutrino intensities is produced.}

The intensity plots of the model pair of jets are created using Veusz, a software for plotting data written by Jeremy Sanders and contributors, and distributed under the GNU/GPL licence. RLOS2 and NEMISS were written by the author and are  available under the lGPL licence. PLUTO was written by Andrea Mignone and collaborators, and is available under GNU/GPL.

{Table \ref{Table-for-run-data} shows a number of simulation parameters. Those include  computational cell length,  jet density, and both winds' maximal densities (those  gradually declined away from their sources). In PLUTO, the piecewise linear method was set up using the MUSCL Hanckock integrator. An ideal equation of state was used. The binary companion is located outside the grid, and was estimated to be at most up to an order of magnitude greater than that of the compact object.  Jet speed is 0.866c, while its kinetic luminosity is \mbox{2.5 $\times$ 10$^{38}$ erg/s}. As a first use of the programme suite, a rather low spatial resolution of 60 $\times$ 100 $\times$ 50 was employed in PLUTO in order to accommodate for the heavier neutrino emission calculation later. }
\clearpage
\startlandscape

\
\begin{table}[H]
\caption{ Five different imaging runs based on  same underlying hydrocode run.}
\label{Table-for-run-data}
\setlength{\cellWidtha}{\columnwidth/7-2\tabcolsep+0.0in}
\setlength{\cellWidthb}{\columnwidth/7-2\tabcolsep+0.0in}
\setlength{\cellWidthc}{\columnwidth/7-2\tabcolsep+0.0in}
\setlength{\cellWidthd}{\columnwidth/7-2\tabcolsep+0.0in}
\setlength{\cellWidthe}{\columnwidth/7-2\tabcolsep+0.0in}
\setlength{\cellWidthf}{\columnwidth/7-2\tabcolsep-0.2in}
\setlength{\cellWidthg}{\columnwidth/7-2\tabcolsep+0.2in}
\scalebox{1}[1]{\begin{tabularx}{\columnwidth}{>{\PreserveBackslash\raggedright}m{\cellWidtha}>{\PreserveBackslash\raggedright}m{\cellWidthb}>{\PreserveBackslash\raggedright}m{\cellWidthc}>{\PreserveBackslash\raggedright}m{\cellWidthd}>{\PreserveBackslash\raggedright}m{\cellWidthe}>{\PreserveBackslash\raggedright}m{\cellWidthf}>{\PreserveBackslash\raggedright}m{\cellWidthg}}
\toprule
\multicolumn{1}{l}{\textbf{Viewing Angle}} & \multicolumn{1}{l}{ \textbf{0 deg}} & \multicolumn{1}{l}{\textbf{10 deg}} & \multicolumn{1}{l}{\textbf{30 deg}}  & \multicolumn{1}{l}{\textbf{60 deg}} & \multicolumn{1}{l}{\textbf{\boldmath{$\simeq$}90 deg}} & \multicolumn{1}{l}{ \textbf{Comments}} \\
\midrule
\multicolumn{1}{l}{$l_{\mathrm{cell}}$ ($\times 10 ^{10}$ cm)} & 2.0 & 2.0 & 2.0 & 2.0 & 2.0 & PLUTO cell \\
\multicolumn{1}{l}{$\rho_{jet}$ (cm$^{-3}$)} & $1.0 \times 10^{10}$ & $1.0 \times 10^{10}$  & $1.0 \times 10^{10}$ & $1.0 \times 10^{10}$ & $1.0 \times 10^{10}$ & Jet matter density \\
\multicolumn{1}{l}{ $\rho_{w}$ (cm$^{-3}$)} & $1.0 \times 10^{13}$ & $1.0 \times 10^{13}$ & $1.0 \times 10^{13}$ & $1.0 \times 10^{13}$ & $1.0 \times 10^{13}$ & Max wind density \\
\midrule
\multicolumn{1}{l}{ $\rho_{dw}$ (cm$^{-3}$)} & $2.0 \times 10^{13}$ & $2.0 \times 10^{13}$ & $2.0 \times 10^{13}$ & $2.0 \times 10^{13}$ & $2.0 \times 10^{13}$ & Max disk wind density  \\
\multicolumn{1}{l}{$t^{max}_{run}$ (s)} & 842  & 842 & 842 & 842 & 842 & Model run time \\
\multicolumn{1}{l}{Method} & P. L.~&  P. L.~& P. L.~&  P. L.~& P. L.~& Piecewise linear \\
\multicolumn{1}{l}{Integrator} & M. H.~& M. H.~& M. H.~& M. H.~& M. H.~& MUSCL-Hancock \\
\multicolumn{1}{l}{EOS} & Ideal &  Ideal & Ideal & Ideal &  Ideal & Equation of state\\
\midrule
\multicolumn{1}{l}{BinSep (cm)} & $4.0 \times 10^{12}$  & $4.0 \times 10^{12}$   & $4.0 \times 10^{12}$  & $4.0 \times 10^{12}$  & $4.0 \times 10^{12}$ & Binary separation \\
\multicolumn{1}{l}{ $M_{BH}/M_{\odot}$} & 3--10 & 3--10 & 3--10 & 3--10 & 3--10 & VE compact star mass \\
\multicolumn{1}{l}{ $M_{\star}/M_{\odot}$} & 10--30 & 10--30 & 10--30 & 10--30 & 10--30 & Companion mass \\
\multicolumn{1}{l}{ $\beta = v_{0}/c$ } & 0.866 & 0.866 & 0.866 & 0.866 & 0.866 & Initial jet speed \\
\multicolumn{1}{l}{ $L^{p}_{k}$} & $ 2.5 \times 10^{38}$ & $ 2.5 \times 10^{38}$ & $ 2.5 \times 10^{38}$ & $ 2.5 \times 10^{38}$ & $2.5 \times 10^{38}$ & Jet kinetic luminosity \\
\multicolumn{1}{l}{Grid resolution} &  60 $\times$  100 $\times$ 50 &  60 $\times$ 100 $\times$ 50 & 60 $\times$ 100 $\times$ 50 &  60 $\times$ 100 $\times$ 50 & 60 $\times$ 100 $\times$ 50 & PLUTO grid size (xyz) \\
\midrule
\multicolumn{1}{l}{Imaging method} &  FB &  PR & PR & PR & FB & \textls[-15]{Focused beam/parallel~rays} \\
\multicolumn{1}{l}{Time delay} &  off &  off & off & off & off & Very high LOS speed\\
\multicolumn{1}{l}{Imaging plane} & XZ-screen & XZ/YZ & XZ/YZ & XZ/YZ & YZ-screen & Box side or inner screen \\
\bottomrule
\end{tabularx}}
\end{table}
\finishlandscape




{ As far as RLOS2 is concerned (synthetic imaging), either focused beam or parallel rays are employed as an imaging method, while the time-delay effect of RLOS2 is not employed at this stage, as it requires multiple RAM memory to be used properly. The synthetic image is projected either on a side of the computational box, either front or sidereal, or on a fiducial imaging screen, again either frontal or sidereal. More specifically, } a series of imaging geometries were employed {following the RLOS2 programme convention}: {(imaging geometry) Case 1,}  parallel rays projected onto the XZ plane;  {Case 2,} the same but onto the YZ plane; {Case 3},  focused rays onto the XZ plane; and {Case 4,} focused rays onto the YZ plane (see also Table \ref{Table-for-run-data}). Three different angles were employed for Cases 1 and 2, while for Cases 3 and 4,  respective focal points implied near-head-on and sidereal views.


RLOS2 \cite{RLOS} was then run using the combined PLUTO--NEMISS data with\mbox{ sfactor = 1} for the pload shrink factor. In general, the imaging process may or may not use all snapshots available to it depending on the light crossing time of its model segment (adjusted through the clight parameter in RLOS2). Trying to read more snapshots than what is loaded corrupts the hydrocode time array of RLOS2, called T, resulting in errors. For simplicity, in our case, an artificially very high clight was used in order to effectively switch off the time-delay effect.  A double filter was used for velocity and for los,u angle. A minimal velocity and  maximal angle were set in order to trigger the calculation of the neutrino emission for a particular cell. This way, the expensive part of the simulation was only performed where it was really worth it. This partly alleviated   the discrepancy between computational costs of the dynamic and the radiative parts of the model.

{The twin jet simulation used in this work represents a single fiducial microquasar using characteristic properties. This system was  dynamically set up to be relatively close to a number of microquasars, such as Cyg X-1 or GRS1915 + 105. From this point on, the model system is imaged with different methods and at different angles in order to explore the perhaps dominant effects of orientation, both locally and globally in the jet. Those imaging results can then be extrapolated to a variety of similar microquasar systems. }

An important aspect of this modelling approach is that each cell has  different visible emissivity  from Earth than that of its neighbours. That is because each cell may differ from the next one in terms of both speed and orientation to us. This combination means that the hydromodel generally gives different results than those of the steady-state one. A vortex with relativistic velocities, for example, may  partly appear very luminous where it is fast with local speed pointing towards us, and also too dark where velocities point away from us. In this simulation, such effects were limited, but at a higher resolution, it is expected that nonlinear dynamic effects in the hydrocode profoundly interact  with the radiative part of the model.

\begin{figure}[H]
\includegraphics[width=10.5 cm]{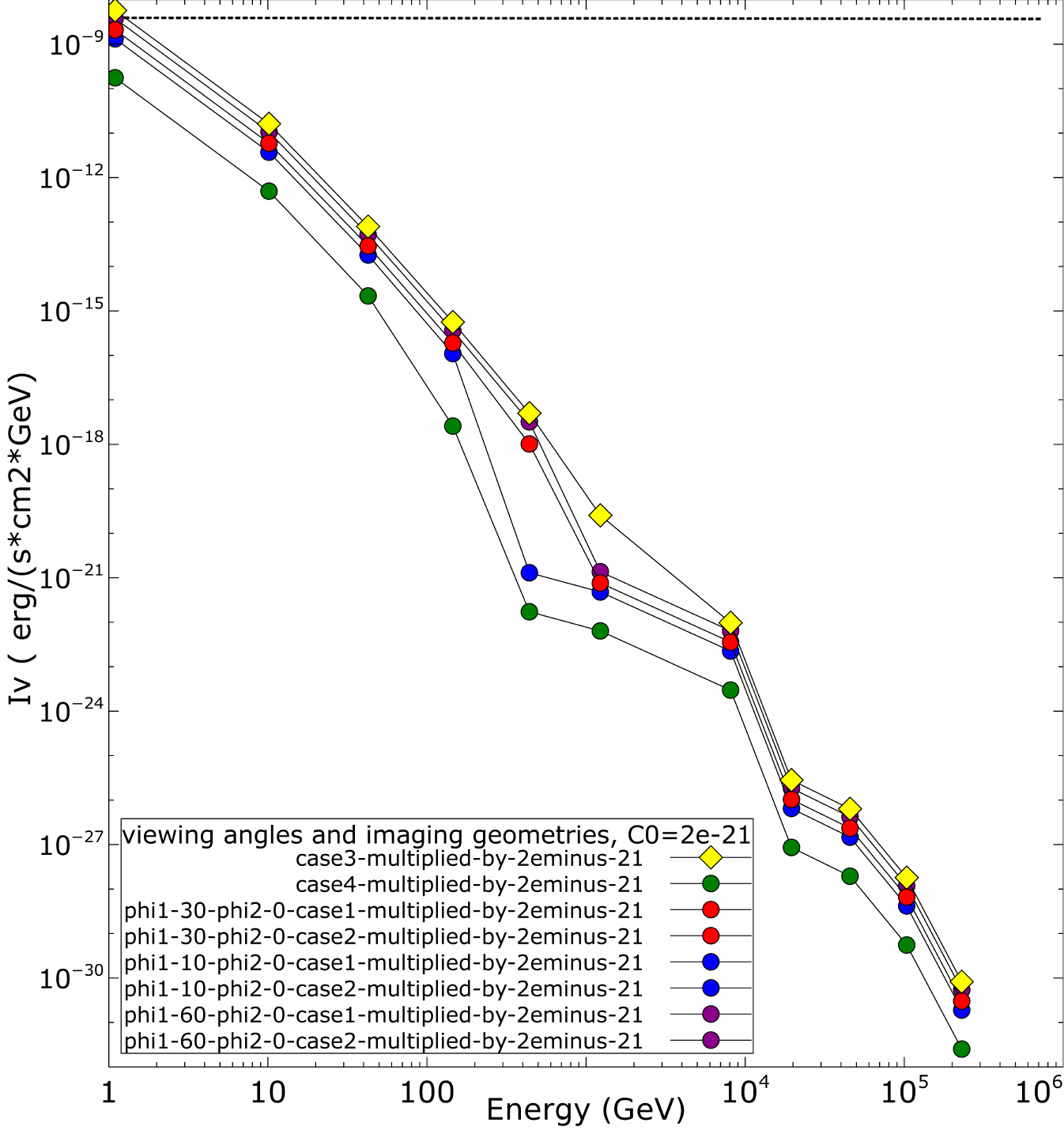}
\caption{Normalized SED from a series of radiative simulations where comparisons with potential observations are  possible. See Appendix \ref{Normalization} for more details on the normalization process. Angle $\phi_{1}$ is complementary to the angle between  jet and the LOS, $\theta$ ($\phi_{1}$ = 90 $-$ $\theta$). Then, { angle} $\phi_{1}$ is nearly 90~degrees when looking along the jet axis. Therefore, $\phi_{1} \simeq$ 90 degrees in imaging geometry's Case 3 {(rays near parallel to the jet axis),} and $\phi_{1} \simeq$ 0 in imaging geometry's Case 4 {(rays nearly perpendicular to the jet axis)}. Intensity decreases as the angle of observation moves away from the jet pair axis. {For comparison, the KM3NeT threshold \cite{Reynoso2009} for detection is included as a dotted line near the top.} Image produced with Veusz using data produced with PLUTO, RLOS2, and NEMISS.}
\label{sed_norm}
\end{figure}

\begin{figure}[H]
\includegraphics[width=10.5 cm]{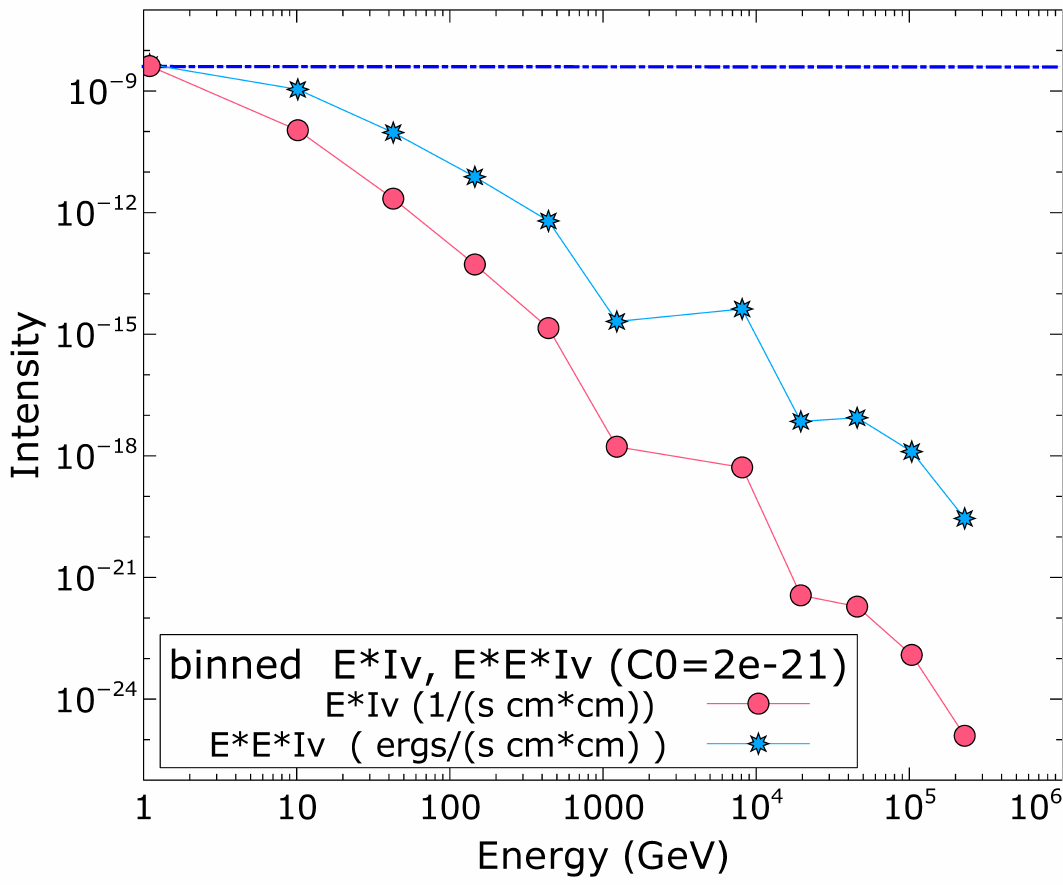}
\caption{ Intensity on Earth at an angle $\phi_{1}$~= 60 degrees, meaning around 30 degrees from the jet axis, weighted by the energy/energy squared of the particle. These represent bin plots, as each data point lies on a higher size scale than the next one. Furthermore, they are also cumulative rate plots, upwards from each given energy, as rate contributions from higher energies are much smaller than the starter one. In comparison with the sensitivity of IceCube{/KM3NeT} of below 10$^{-8}$ in this plot,  results seem marginally acceptable in anticipation of potential detection. Image produced with Veusz. {For comparison, the KM3NeT threshold for detection is included as a dotted line near the top.} }
\label{weighted_plots}
\end{figure}

The scale of the total emission increases the closer that the LOS approaches to the jet pair axis (Figure \ref{sed_norm}). The employed low resolution  did not allow for significant nonlinear dynamic effects to appear, yet the concept of the modelling process was proven to work in its entirety. On the other hand, the normalization process demonstrates the possibility of potential observations, as the results potentially fall within the detection range of contemporary arrays  \cite{Reynoso2009}. {The detection ability of km$^{3}$ array is depicted in the normalized spectral emission distribution (SED) plots, as a measure of comparison with the model results. A certain potential for detection appears that is rather promising to explore.}
More specifically, we can see in Appendix \ref{Normalization} that  intensity on Earth is proportional to  kinetic jet luminosity $L_{k}$ and  inversely proportional to the square of the distance to us $D^{2}$. Consequently, a sample set of rates can be extracted from the model and  used as a reference for other microquasars at different distances and with different jet energies than those of the standard.  Figure \ref{weighted_plots} shows the weighted set of rates expected on Earth for a sample microquasar viewed at 30 degrees from the jet axis with $L_{k}$ = 10$^{38}$ergs$^{-1}$ and \emph{D} = 5 kpc. Other systems  then have $I=I_{0,\theta}\frac{L_{k}}{L_{k0}} (\frac{D_{0}}{D})^{2}$, where \emph{I}, $L_{k}$, and \emph{D} refer to a new microquasar; $I_{0}$, $L_{k0}$, and $D_{0}$ represent the standard plotted here, and the profound effect of the viewing angle is implicitly included. The result of Figure \ref{weighted_plots} is comparable to the sensitivity of {the state-of-the-art instrument arrays IceCube/KM3NeT} {IceCube} in terms of a squared energy weighted curve, which falls below 10$^{-8}$ throughout the plot's energy span \cite{icecube_sens}.

The above estimate may then be employed in order to provide a rough estimate of expected neutrino emission from a distribution of microquasars in the galaxy. The authors of \cite{Paredes2003} argued an estimated population of approximately one-hundred systems in our galaxy. Furthermore, their discussion of $\gamma$ ray emission from microquasars clarifies the importance of relativistic boosting in jet emission. Thus, orientation to Earth plays a major role here, and the situation is similar for neutrino emission.

We proceed by accepting 100 systems at various distances ranging from a minimum of 1 kpc to a maximum of 30 kpc, with  average kinetic luminosity similar to our model system. The linear dependence of emissions on the latter quantity facilitates such a  simplification. A distance of 1 kpc commands a flux at Earth of 25 times more than our model value, whereas a system situated at 30 kpc   has 36 times less than that. Last, an orientation of less than 60 degrees might be 1 order of magnitude less than our value, but a jet system aimed towards us  could have up to 100 times more visibility at Earth unless a very fast jet occurred. Consequently, the single most important factor is orientation, followed by distance and lastly by jet kinetic power. The latter order allows for an estimate of perhaps 5\%, or five systems with a very high relativistic boosting towards us, a number of maybe 40 or 50 at angles above 45 degrees, and lastly maybe 50 at below 45 degrees. The first five  probably contribute the most on average, and the ones viewed from the side  have a smaller effect. A possible system at a smaller distance would of course dominate the distribution, but the possibility for such an occurrence is questionable.

\begin{figure}[H]
\includegraphics[width=10.5 cm]{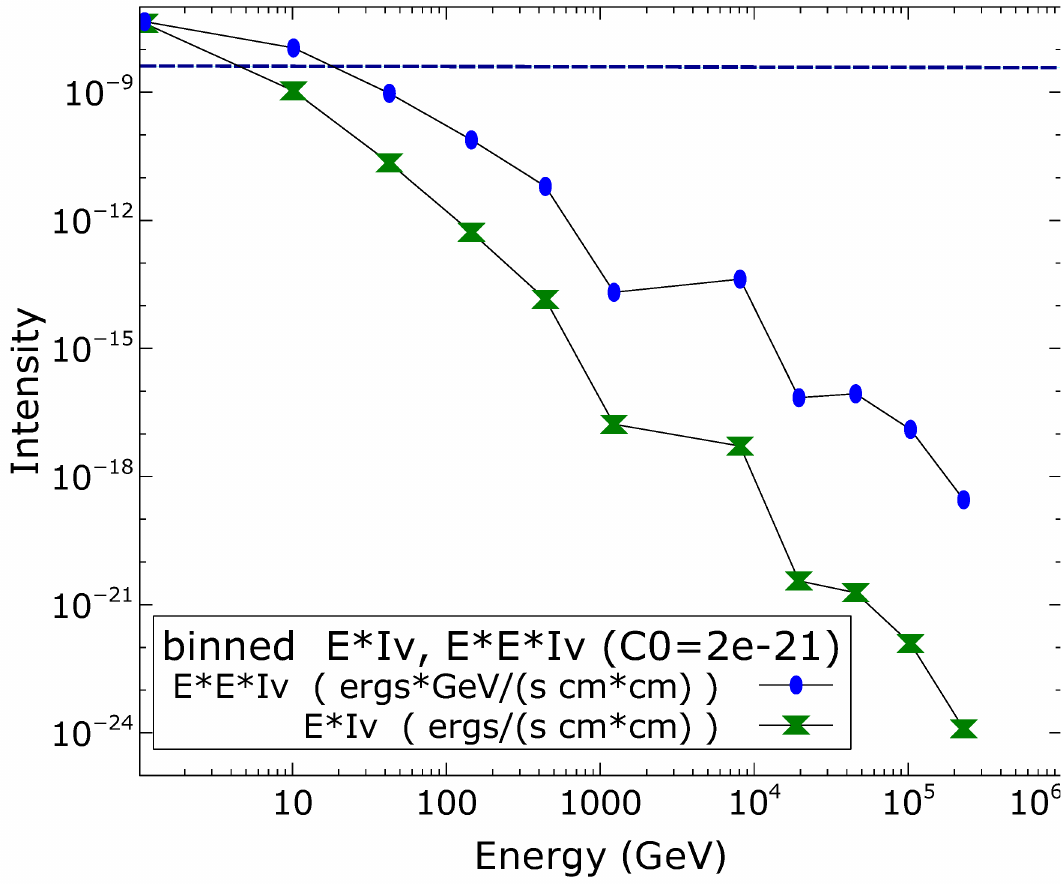}
\caption{ Intensity on Earth, of a fiducial distribution of systems weighted by the energy/energy squared of the particle. Again, these may represent bin plots due to the logarithmic decline of the quantity. For the same reason they are also cumulative rate plots, upwards from each given energy. {For comparison, the KM3NeT threshold for detection is included as a dotted line near the top.} Image produced with Veusz.}
\label{distro_weighted_plots}
\end{figure}

On the basis of the above discussion, we then accept a rough average for a neutrino-emitting galactic microquasar located at 15 kpc, with the kinetic luminosity of our model (less affecting factor) and orientated at 30 degrees from the line of sight, which is the case used in Figure \ref{weighted_plots}. The reason for having the average angle at less than 45 degrees is the higher contribution from systems aimed towards as. We then multiply our single microquasar result by 100 (population size), divide it by 3$^{2}$ (distance) and leave the jet power effect at unity. A rough first estimate could then be to multiply our single system result at 30 degrees from the jet axis by a factor of ten (Figure \ref{distro_weighted_plots}) and then use it for comparison with observations.

{Orientation seems to play a crucial role here and   is thus given the primary role in the synthetic imaging process by employing various orientation scenarios for the model pair of jets. In addition, this model calculates the effects of orientation at each point of the 3D PLUTO twin jet simulation, and then produces a synthetic neutrino image. Thus, the important effects of differential projection effects are explored, paving the road for more detailed simulations in the future using this programme suite. }

{In contrast, previous similar works \cite{SK15,SK17}, calculated neutrino emission at just a handful of points along a single model jet (a much smaller computational task), and then used a semi-analytic approach to cover the rest. Furthermore, a number of programme improvements were  incorporated into the models, such as using \cite{TR11} for relativistic orientation and velocity transform of the hot proton distribution, as opposed to \cite{PS2001} in the previous works where this author contributed.  }

The above results {for  microquasar distribution}  may vary to either direction by possibly an order of magnitude, subject to a more detailed statistical analysis. This is because there are similar systems with higher or lower jet kinetic power, as well as systems with various individual properties. Nevertheless, it seems possible that the detection of a background emission from a potential distribution of microquasars in the galaxy is within the realm of modern detector arrays. This is also a consideration for the next generation of new or upgraded arrays being planned today. On the other hand, a single X-ray binary system also looks promising as a galactic source of high-energy neutrinos. This is a potential target for a particle sensor with increased angular accuracy. The variability of microquasars within the human timescale, combined with their relative stability as a known point source, offers a good target for observation, especially combined with sensors working in electromagnetic spectra, such as radio, X-rays, and $\gamma$ rays. In such a case, a neutrino observation of a microquasar may form part of a multi-wavelength observation effort aimed at the system of~interest.



\vspace{-6pt}

\section{Final Remarks}
\label{conclusions}

Particle emission from a typical microquasar was simulated using a suitable programme suite. Results verified the integrity of the process, paving the way for more detailed runs. Furthermore, the model was employed in order to provide particle emission estimates   for both a single microquasar and a potential galactic distribution of such systems. The latter approach facilitates  a comparison with the output of contemporary detection arrays, where microquasars  could contribute to a background of high-energy neutrinos.

In the model, a series of both dynamical and imaging parameters may be adjusted in order to cover different scenarios. The programme suite works in a highly automated manner, and is prepared to take on higher-resolution applications where the relativistic effects of nonlinear dynamics may appear in full.

The ability to focus on individual cells could greatly differentiate each jet element from the next in terms of emission. An MHD jet has great local variability in both particle and radiation emission intensity in any given direction. The detailed dynamics of the jet  influence the appearance of the system depending on both the direction and  magnitude of the local velocity, and on pressure and  density. Consequently, a jet system with turbulence, vortices, colliding with clouds, etc. is expected to be subject to the aforementioned local variations of intensity.

As far as absorption is concerned, the model may directly include the emission and absorption of electromagnetic radiation at different frequencies. Should adequate computing resources be employed, the time-delayed description in the programme can also be activated. For example, a turbulent relativistic jet colliding with a cloud has different parts of it moving at high velocities in different directions. The image is then dynamically formed, the rays crossing a choreography of relativistically moving jet elements. The final image may be quite different than what is initially expected, as demonstrated, for example, by the effect of apparent superluminal motion. For a complex jet system, running the model at higher resolutions with the time-delay module could reveal many physical details, drawing a more realistic picture of the system.

In general, microquasars may locally emit at reinforced levels of intensity due to the combination of jet dynamics and relativistic projection. The reason can be internal jet turbulence or interaction with clouds and surrounding winds. For $\gamma$ rays and neutrinos, such dynamic effects should occur in the vicinity of the jet base.

{
Furthermore, the  employed model can be used as a basis for expanding the approach to systems of different scale. The innermost AGN jets can be sources of ultra high-energy cosmic rays \cite{Dermer_2009}, and PLUTO can model those jets with a suitable set of initialization parameters. Special relativistic MHD should be employed as an approximation, though. There is a possibility to include a quasi-Newtonian potential as an improved approximation for the innermost part of a quasar  jet. The emission model, which in our case was  NEMISS, should be altered in order to include the new emission physics. Synthetic imaging code RLOS2 is ready to use with any emission and absorption input, and only minor changes are required.}

{ Further out along an AGN jet, neutrino emission may occur from high-energy proton acceleration along with other signals such as $\gamma$ rays  \cite{oikonomou2021}. This description is similar to microquasars, and only the scales differ. Consequently, it should be possible to suitably adapt the current simulations  in order to model neutrino emission from the inner part of a quasar jet. }

\vspace{6pt}

\funding{{This research received no external funding.}}

\institutionalreview{{Not applicable}.}

\informedconsent{{Not applicable}.}

\dataavailability{{Synthetic data used in this paper were produced with code linked to in the bibliography}.}

\acknowledgments{We thank R. E. Spencer (Jodrell Bank Observatory) for the valuable comments on the manuscript. Special thanks go to G. E. Romero (UNLP, IAR) for his suggestions on improving the content of this {work.} }

\conflictsofinterest{The author declares no conflict of interest.} 



\bibliography{biblio_100821}

\appendixtitles{yes} 
\appendixstart
\appendix


\section{{Normalization}}
\label{Normalization}

The {special relativistic } kinetic energy of the jet at its base can be expressed as  {\cite{Reynoso2009} }

\begin{equation}
E_{k}= \frac{1}{2} (\Gamma m) u^{2}
\end{equation}
\textls[-5]{where \emph{u} is  jet speed, and m the mass of a jet portion crossing the cross section of the jet there. Then,  jet kinetic power $P_{k}$ is the kinetic energy traversing the cross section per unit time}
\begin{equation}
P_{k}=dE_{k}/dt= \frac{1}{2} (\Gamma dm/dt) u^{2}
\end{equation}
where the speed is taken to be constant during an ejection episode (it was also set to be constant in the simulation described here). However,
\begin{equation}
dm/dt=\rho dV/dt = \rho A dx/dt =\rho A u
\end{equation}
where \emph{A} is the jet base cross section area, also taken  as a constant both in the simulation and here. The volume element \emph{dV} equals $Adx$. Therefore,
\begin{equation}
P_{k}=dE_{k}/dt= \frac{1}{2} (\Gamma \rho A) u^{3}
\end{equation}
or
\begin{equation}
P_{k}=dE_{k}/dt= \frac{1}{2} (\Gamma \rho N_{\mathrm{cell}} L^{2}_{\mathrm{cell}}) u^{3}
\end{equation}
where $L$$_{\mathrm{cell}}$ is the cell length, $N$$_{\mathrm{cell}}$ is the number of cells forming the cross section of the jet base, and $A=N_{\mathrm{cell}} L^{2}_{\mathrm{cell}}$. The square $L^{2}_{\mathrm{cell}}$ is then the area of the side of length $L_{\mathrm{cell}}$ of a cubical computational cell at the jet base. We then express density as a function of proton number density $N_{p}$ and proton mass $m_{p}$,
\begin{equation}
\rho=N_{p} m_{p}.
\end{equation}

{ Let us define  neutrino luminosity $L_{\nu}$ as the power emitted through neutrinos from the jet, which is a fraction $\alpha$ of the total kinetic jet power (jet kinetic luminosity $L_{k}$),}  $L_{\nu}=\alpha L_{k}$, representing the portion of jet power emitted in neutrinos. For  normalisation, a working value is taken as 10$^{-3}$. This can be justified from a q$_{\mathrm{rel}}$ = 0.1 for the energy content of the relativistic particles in the jet \cite{Reynoso2009,Reynoso2019}, on top of which we employ the efficiency of the cascade when transferring energy from hot protons to final neutrinos.

The shape of the spectrum is also affected by  acceleration efficiency \cite{Reynoso2019}, and from the opening angle of the jet \cite{Reynoso2009}, thus affecting the area under the neutrino spectrum plot. As an approximation for the above effects, we adopted a value of 0.01 for the energy transfer from non-thermal protons to the neutrinos.

{Furthermore, we introduce a factor $\alpha=L_{\nu}/L_{k}$, representing the portion of jet power emitted in neutrinos. A typical value is taken as 10$^{-3}$.}
We also set u = $\beta$c. A less-than-unity positive filtering factor $f_{f}$ is employed that  accounts for not using all   jet cells, but only those with velocity orientation closer to the LOS and  with speed above a given limit.
We then have
\begin{equation}
L_{\nu}=\alpha L_{k}=\alpha P_{k}=\alpha dE_{k}/dt= f_{f} \frac{1}{2} \alpha \Gamma (N_{p} m_{p} N_{\mathrm{cell}} L^{2}_{\mathrm{cell}}) \beta^{3} c^{3}
\end{equation}

The intensity of the jet is then expressed as $I_{\nu}=L_{\nu}/4 \pi D^{2}$, where \emph{D} is the distance to Earth. Thus,
\begin{equation}
I_{\nu}= f_{f} \frac{1}{4 \pi D^{2}} \alpha \frac{1}{2} \Gamma (N_{p} m_{p} N_{\mathrm{cell}} L^{2} _{\mathrm{cell}}) \beta^{3} c^{3}
\end{equation}

\textls[-15]{In our simulation, the jet beam travels at $\beta$ = $\frac{u}{c} = $0.866, with a density of 10$^{10}$ protons/cm$^{3}$}. $L_{\mathrm{cell}}$ is 10$^{10}$ cm, while the number of cells comprising the beam at its base at this resolution is $N_{\mathrm{cell}} \simeq$ 15. Furthermore, $\Gamma$~=~2. Distance to Earth is taken here with a typical value of \emph{D} = 5 kpc or approximately 2$\times$ 10$^{22}$ cm.
We then integrate the area under the curve of an arbitrary units neutrino intensity plot, for the case of nearly non-beamed data, at $\phi_{1}$~=~10 degrees. That case is supposed, for the purposes of normalization, to be the one matching the orientation of the hypothetical system in relation to Earth. We perform a cumulative sum over the roughly 10 points, admitting  10\% coverage per order of magnitude scale level. Thus, we find about 10$^{11}$, which means that our sum is 10 times smaller, or approximately 10$^{10}$, in (AU)*GeV, where AU stands for arbitrary units. We replace an AU with a constant \emph{C}$_{0}$, so that AU~= \emph{C}$_{0}$ erg/(s*cm$^{2}$).
We set $I_{\nu}$~=~$L_{\nu}/4 \pi D^{2}$ equal to the area under the un-normalized intensity plot with neutrino energy, expressed in units of \emph{C}$_{0}$, in order to find the latter (normalization constant)
\begin{equation}
I_{\nu}= f_{f} \frac{1}{4 \pi D^{2}} \alpha \frac{1}{2} (\Gamma (N_{p} m_{p}) N_{\mathrm{cell}} L^{2} _{\mathrm{cell}}) \beta^{3} c^{3}= (\mathrm{PLOT AREA})*C_{0} \, \mathrm{erg/(s \, cm^{2}) \, GeV}
\end{equation}

For our case, we find $C_{0}$ $\simeq$ 2~$\times10 ^{-21}$, which is the value of the arbitrary unit $C_{0}$. Using the above constant, we multiply by it the value given in arbitrary units for the particle emission.  Thus, the intensity plot is multiplied, and we arrive to the updated plot in \mbox{Figure~\ref{sed_norm}}, which may be directly compared to other models and to observations.
\section{{Equipartition Calculation} 
}
\label{Equipartition_calculation}

The equipartition calculation now follows. As shown above, the jet kinetic power is
\begin{equation}
\L_{k}=\frac{1}{2}\frac{dm}{dt} u^{2}=\frac{1}{2} (\rho A u) u^{2}=\frac{1}{2} \rho A u^{3}
\end{equation}
where $\frac{dm}{dt}=\rho \frac{dV}{dt}=\rho A \frac{dx}{dt} =\rho A u$

Kinetic energy density, of a jet with radius R$_{j}$, is \cite{Reynoso2009}
\begin{equation}
\rho_{k}=\frac{L_{k}}{\pi R_{j}^{2} u_{j}}=\frac{L_{k}}{A u}= \frac{\frac{1}{2} \rho A u^{3}}{A u} =\frac{1}{2} \rho u^{2}
,\end{equation}
which also acts as verification.

We also have for the local magnetic field
\begin{equation}
B=\sqrt{8 \pi \rho_{B}}
\end{equation}

For equipartition, we set the kinetic and magnetic energy densities to be equal to each other, $\rho_{k}=\rho_{B}$. Therefore,
\begin{equation}
B=\sqrt{8 \pi \rho_{B}}=\sqrt{8 \pi \rho_{k}}
\end{equation}

We  now have $\rho_{k}$;  then, we  calculate the \emph{B} that corresponds to equipartition for that. Our beam has $\rho$~=~10$^{10}$~cm$^{-3}$, or about 1.6 $\times$ 10$^{-14}$~gcm$^{-3}$. Thus,
\begin{equation}
\rho_{k}=\frac{1}{2}(1.6 \times 10^{-14}\mathrm {\frac{g}{cm^{3}}})(\frac{\sqrt{3}}{2} 3 \times 10^{10} \mathrm{ \frac{cm}{s} } )^{2}\simeq 5.4 \times 10^{6} \mathrm{\frac{g}{\mathrm{cm s^{2}}}}
\end{equation}

Therefore, in CGS
\begin{equation}
B=\sqrt{8 \pi \rho_{k}}\simeq\sqrt{8 \pi ~ 5.4\times 10^{6}}\simeq 11.5 \times 10^{3}
\end{equation}

For our simulation we set a rounded value of \emph{B}~=~10$^{4}$ G, which is not far from the approximate equipartition value found above.


\reftitle{References}

\end{document}